\documentstyle[12pt,epsf,epsfig,wrapfig]{article}
\begin{document}

\begin{center}
{\bfseries Chiral filtering of spin states as a source of SSA}

\vskip 5mm
\underline{S.M. Troshin$^\dag$} and
N.E. Tyurin

\vskip 5mm
{\small
 {\it
Institute for High Energy Physics, Protvino, Russia
}\\
 {\it
$^\dag$E-mail: Sergey.Troshin@ihep.ru
}}
\end{center}

\vskip 5mm
\begin{abstract}
We discuss simple nonperturbative mechanism for generation of single-spin
asymmetries in hadron processes and hyperon polarization in particular.
\end{abstract}
\vskip 8mm
The perturbative QCD is well elaborated
theory and
provides predictions for many phenomena and among them are helicity conservation
in hard processes and existence of the new form of matter with free quarks and gluons
-- quark-gluon plasma.  However, violation
of helicity conservation has been observed in many experiments and has a rather long
 experimental history,
while strongly interacting
deconfined  matter in heavy-ion collisions at RHIC was revealed just recently.
It appeared that instead of gas of free quarks and gluons the matter created at RHIC is an almost
perfect liquid. The nature of this new form of matter is not known and the variety of models
has been proposed to treat its properties. The importance of this
result is that beyond the critical values of density and temperature
 the matter is strongly correlated
and reveals high degree of the coherence. The elliptic flow and constituent quark scaling
 demonstrated  an importance of the constituent quarks and their role as
effective degrees of freedom of the newly discovered form of matter.
Generally speaking this result has shown again an importance of the nonperturbative
effects in the region where such effects were usually not expected.

One of the most interesting and persistent for a long time spin phenomena
was observed in inclusive hyperon production in collisions of
unpolarized hadron beams. A very significant polarization of
$\Lambda$--hyperons has been discovered almost three  decades ago\cite{newrev}.
  Experimentally the process
of $\Lambda$-production has been studied more extensively than other hyperon
production processes.
Therefore we will emphasize on the particular riddle  of $\Lambda$--polarization
because  spin structure of this particle is most simple and
 is determined by strange quark only. This mechanism can also
be used for the explanation of single-spin asymmetries in the inclusive pion production.

It should be noted that understanding of
transverse single-spin asymmetries in DIS (in contrast to the hyperon polarization)
has observed significant progress during last years; this progress is related
to an account of final-state interactions from gluon exchange\cite{brodsky} --
coherent effect not suppressed in the Bjorken limit.

Experimental
 situation with hyperon polarization is widely known and stable for a long time.
Polarization of $\Lambda$ produced in the unpolarized inclusive $pp$--interactions
is negative and energy
independent. It increases linearly with $x_F$ at large transverse momenta
($p_\perp\geq 1$ GeV/c),
and for such
values of transverse momenta   is almost
$p_\perp$-independent\cite{newrev}.

On the theoretical side,  perturbative QCD
with a straightforward collinear factorization scheme
leads to small values of $\Lambda$--polarization
which are far below of the corresponding experimental data.
Modifications of this scheme and  account for higher twists contributions allows
to obtain higher magnitudes of polarization but do not change
a decreasing  dependence proportional to
$p_\perp^{-1}$ at large transverse momenta.
It is difficult to reconcile this behavior  with the flat  experimental
data dependence on the transverse momenta. Inclusion of the internal transverse momentum
of partons ($k_\perp$--effects) into the
polarizing fragmentation functions  leads   to similarly decreasing polarization\cite{anselm}.
In addition it should be noted that the perturbative QCD has also problems in the description
of the unpolarized scattering, e.g. in inclusive cross-section
for $\pi^0$-production, at the energies lower than the RHIC energies\cite{bsof}.

The essential point  of the approaches mentioned above is that the vacuum at short distances
is taken to be a perturbative one.
There is an another possibility. It might happen  that
the polarization dynamics in strangeness production originates from the genuine
nonperturbative sector of QCD.
In the nonperturbative sector of QCD the  two important
phenomena,  confinement and spontaneous breaking of chiral symmetry ($\chi$SB)
should be reproduced.
The  relevant scales   are characterized by the
parameters $\Lambda _{QCD}$ and $\Lambda _\chi $, respectively.  Chiral $SU(3)_L\times
SU(3)_R$ symmetry is spontaneously  broken  at the distances
in  the range between
these two scales.  The $\chi$SB mechanism leads
to generation of quark masses and appearance of quark condensates. It describes
transition of current into  constituent quarks.
  Constituent quarks are the quasiparticles, i.e. they
are a coherent superposition of bare  quarks, their masses
have a magnitude comparable to  a hadron mass scale.  Therefore
hadron  is often represented as a loosely bounded system of the
constituent quarks.
These observations on the hadron structure lead
to  understanding of several regularities observed in hadron
interactions at large distances. It is well known  that such picture  provides
reasonable  values  for the static characteristics of hadrons, for
 instance, their magnetic moments. The other well known direct result
   is  appearance of the Goldstone bosons.

Recently nonperturbative approach  to SSA
 has been developed in\cite{burk}.
It is based
on the parton picture in the impact parameter space and assumed specific helicity--flip
generalized parton distribution.
The instanton--induced mechanism of SSA generation was considered in\cite{koch,shur}
and relates those asymmetries with a genuine nonperturbative QCD interaction.

As it was already mentioned constituent quarks and Goldstone bosons are the effective
degrees of freedom in the chiral quark model. We consider a
 hadron consisting of the valence
constituent quarks located in the central core which is embedded into  a quark
condensate. Collective excitations of the condensate are the Goldstone bosons
and the constituent quarks interact via exchange
of Goldstone bosons; this interaction is mainly due to a pion field which is of the flavor--
 and spin--exchange nature. Thus, quarks generate a strong field which
binds them\cite{diak}.

At the first stage of hadron interaction common effective
self-consistent field is appeared.
Valence constituent quarks   are
 scattered simultaneously (due to strong coupling with Goldstone bosons)
and in a quasi-independent way by this effective strong
 field. Such ideas were already used in the model\cite{csn} which has
been applied to description of elastic scattering and hadron production\cite{mult}.

The initial state  particles (protons) are unpolarized.
It means that states with spin up and spin down have equal probabilities.
The main idea of the proposed mechanism is the  filtering
of the two initial spin states of equal probability due to different strength of interactions. The particular
mechanism of such filtering can be developed on the basis of chiral quark model,
formulas for inclusive cross section (with account for the unitarity)\cite{tmf} and
notion on the quasi-independent nature of valence quark scattering in the effective field.

We will exploit the feature of chiral quark model that constituent quark $Q_\uparrow$
with transverse spin in up-direction can fluctuate into Goldstone boson and
  another constituent quark $Q'_\downarrow$ with opposite spin direction,
   i. e. perform a spin-flip transition\cite{cheng}:
\begin{equation}\label{trans}
Q_\uparrow\to GB+Q'_\downarrow\to Q+\bar Q'+Q'_\downarrow.
\end{equation}
An absence of arrows means that the corresponding quark is unpolarized.
To compensate quark spin flip $\delta {\bf S}$ an orbital angular momentum
$\delta {\bf L}=-\delta {\bf S}$ should be generated in final state of reaction (\ref{trans}).
The presence of this orbital momentum $\delta {\bf L}$  in its turn
means  shift in the impact parameter
value of the final quark $Q'_\downarrow$ (which is transmitted to the shift in the impact
parameter of $\Lambda$)
\[
\delta {\bf S}\Rightarrow\delta {\bf L}\Rightarrow\delta\tilde{\bf b}.
\]
Due to   different strengths of interaction at the different values of the
impact parameter, the processes of transition to the
spin up and down states will have different probabilities which  leads eventually to
polarization of $\Lambda$.

It is important to note here that the shift of $\tilde{\bf b}$
(the impact parameter of final hyperon)
is translated to the shift of the impact parameter of the initial particles according
to the relation between impact parameters in the multiparticle production:
\begin{equation}\label{bi}
{\bf b}=\sum_i x_i{ \tilde{\bf  b}_i}.
\end{equation}
The variable $\tilde b$ is conjugated to the transverse momentum of $\Lambda$,
but relations  between functions depending on the impact parameters
$\tilde b_i$ are nonlinear.

We consider production of $\Lambda$ in the fragmentation region, i.e.
at large $x_F$ and therefore use approximate relation
$b\simeq x_F\tilde b$,
which results from Eq. (\ref{bi}).

The mechanism of the polarization generation  is quite natural
 and has an optical analogy with the passing
 of the unpolarized light through the glass of polaroid.
The particular mechanism of filtering of spin states is related to the
 emission of Goldstone bosons by constituent quarks.

In a particular case of $\Lambda$--polarization the relevant transitions
of constituent quark $U$  will be correlated with the shifts $\delta\tilde b$
in impact parameter $\tilde b$ of the final
$\Lambda$-hyperon, i.e.:
\begin{eqnarray}
  \nonumber U_\uparrow & \to & K^+ + S_\downarrow\Rightarrow\;\;-\delta\tilde{\bf b} \\
\label{spinflip} U_\downarrow & \to & K^+ + S_\uparrow\Rightarrow\;\;+\delta\tilde {\bf b}.
\end{eqnarray}
Eqs. (\ref{spinflip}) clarify mechanism of the filtering of spin states:
 when shift in impact
parameter is $-\delta\tilde {\bf b}$ the
interaction is stronger compared to the case when shift is $+\delta\tilde {\bf b}$,
and the final $S$-quark
(and $\Lambda$-hyperon) is polarized negatively.
Eqs. (\ref{spinflip}) clarify mechanism of the filtering of spin states:
 when shift in impact
parameter is $-\delta\tilde {\bf b}$ the
interaction is stronger compared to the case when shift is $+\delta\tilde {\bf b}$,
and the final $S$-quark
(and $\Lambda$-hyperon) is polarized negatively.

The explicit formulas for inclusive
cross--sections of the process
 $h_1 +h_2\rightarrow h_3^\uparrow +X$, where hadron $h_3$ is a hyperon whose
transverse polarization is measured were obtained in
\cite{tmf}. The main feature of this formalism is an account of
unitarity in the direct channel of
reaction.
Together with unitarity, which is an essential
ingredient of this approach, the filtering mechanism allows  to
 obtain results for polarization dependence on kinematical variables
 in  agreement with the  experimental  behavior
of $\Lambda$-hyperon polarization, i.e.
 linear dependence on $x_F $ and
flat dependence
on $p_\perp$ at large $p_\perp$
in the fragmentation region are reproduced.
Those dependencies together with the energy independent
behavior of polarization at large transverse
momenta are the straightforward consequences of this model.

We discuss here  particle production in the fragmentation region.
In the central region where correlations
 between impact parameter of the initial and impact parameters of the final particles
 being weakened, the polarization cannot be generated due to chiral quark filtering
 mechanism.
Moreover, it is  clear that since antiquarks are produced through spin-zero Goldstone bosons
we should expect $P_{\bar\Lambda}\simeq 0$.
The chiral quark filtering is also relatively suppressed when compared to direct elastic
 scattering of quarks in effective   field and therefore
   should not play a role in the reaction $pp\to pX$ in the fragmentation
 region, i.e. protons should be produced unpolarized. These features take place
 in the experimental data set.

SSA is an interesting topic not only in the field of hyperon polarization.
The new experimental
expectations are related to the experiments at RHIC with polarized proton beams and new
experimental data obtained by the STAR collaboration have already demonstrated significant
spin asymmetry in the $\pi^0$--production similar to the one observed
in the fragmentation region at FNAL.

We would like to make a brief comment on this subject and to
note that the reverse to the filtering mechanism can be used for the
 explanation of the SSA in pion
production observed at FNAL and recently at RHIC in the fragmentation region.
In the initial state of these reaction the proton is polarized
 and can be represented in the simple SU(6) model as  following:
 \begin{equation}\label{pr}
 p_\uparrow=\frac{5}{3}U_\uparrow+\frac{1}{3}U_\downarrow+\frac{1}{3}D_\uparrow+
 \frac{2}{3}D_\downarrow.
\end{equation}
The relevant process for $\pi^+$--production is
\[
U_\uparrow\to\pi^+ + D_\downarrow,
\]
which leads to a negative shift in the impact parameter and consequently
to the positive asymmetry $A_N$, while the corresponding process for
the $\pi^-$--production
\[
D_\downarrow\to\pi^- + U_\uparrow .
\]
It leads to the positive shift in impact parameter and respectively
 to the negative asymmetry $A_N$.
Asymmetry $A_N$ in the fragmentation region should have similar to polarization linear $x_F$--dependence which
 is in agreement with the observed FNAL and RHIC experimental data.
  As for the neutral
  $\pi^0$--production the combination of $U$ and $D$--quarks with up and down polarization
  makes contributions to cross--sections and asymmetry. On the basis of the simple SU(6)
model we can assume that the $U$--quark with up polarization would contribute mainly
in the fragmentation region. Then the $\pi^0$--production should have positive asymmetry.
 Linear $x_F$-dependence agrees with the experimental data at large $x_F$.

It should be noted that the unpolarized inclusive
cross-section of the pion production is proportional to $p_\perp^{-6}$
dependence at large $p_\perp$ and  is valid also
for the pion production. It is in  agreement
with $p_\perp^{-N}$ (with the exponent $N=6.2\pm 0.6$) dependence of
the inclusive cross-section of $\pi^0$-production observed
 in forward region at large $p_\perp$ at RHIC \cite{star}.

The proposed mechanism deals with effective degrees of freedom and takes into
account collective aspects of the nonperturbative QCD dynamics.
We discussed here  particle production in the fragmentation region and have
shown that the power-like behavior of the differential cross-sections at large transverse
momenta can be obtained in the approach which has a nonperturbative origin.
 It might happen that
the transient stage in hadron and heavy-ion interactions have
a common nature and represent a strongly
interacting matter of constituent quarks.
This interaction should then be identical to the one
responsible for  formation of more
 complex multiquark states. Such interaction would result in collective
 rotation of the quark matter as a whole produced in the non-central heavy-ion collisions.
 This rotation should be taken then into account since it will affect the elliptic flow and
 would probably result in strong spin correlations of the produced particles similar to
  correlations predicted for hadron processes \cite{ipl}.

 We would also like to note that
this model explains an exponential dependence of inclusive
cross--section in the region of moderate transverse momenta and
  provides a reasonable description of
the energy dependence of mean multiplicity leading to
its power-like growth with a small exponent \cite{jpg} as
a combined effect of the unitarity and account for the  phase preceding
 hadronization when massive quark--antiquark pairs are generated.

\end{document}